\documentclass[prl,aps,twocolumn,showpacs]{revtex4}
\usepackage{graphicx}
\usepackage{amsmath,amssymb,revsymb}
\begin{document}

\title{Efimov Trimer Formation via Ultracold Four-body Recombination}
\author{Yujun Wang}
\author{B.D. Esry}
\affiliation{Department of Physics, Kansas State University, Manhattan, Kansas 66506, USA}

\begin{abstract}
We discuss the collisional formation of Efimov trimers via ultracold four-body recombination.
In particular, we consider the reaction $A$+$A$+$A$+$B$$\rightarrow$$A_3$+$B$ with
$A$ and $B$ ultracold atoms.  We obtain expressions for the {\em four}-body recombination rate and
show that it reflects the {\em three}-body Efimov physics either as a function of
collision energy or as a function of the two-body $s$-wave scattering length between $A$
atoms.  In addition, we briefly discuss issues important for experimentally observing
this interesting and relatively unexplored process.
\end{abstract}

\maketitle

The exquisite experimental control possible in ultracold atomic gases has enabled
the observation of unique and bizarre quantum states. 
For instance, the many-body phenomena of Bose-Einstein condensation~\cite{JILAObs,MITObs}
and Fermi degeneracy~\cite{HaraFermi,OhashiCross} have both been 
observed and used for a wide range of studies~\cite{Vortex,Mott} 
Weakly-bound diatomic Feshbach ``halo'' molecules have also been observed~\cite{FeshbachMol}
and their dynamics probed~\cite{GrimmHalo}. 
Especially relevant to the present work, the Efimov effect~\cite{Efimov} 
was finally confirmed experimentally via three-body recombination of cesium atoms~\cite{Efimov_Exp}. 
This experimental observation of Efimov physics was preceded by a considerable amount 
of theoretical work due to its general importance for ultracold three-body 
collisions (see, for instance, Refs.~\cite{MacekRecomb,EsryRecomb,Jose_Scale,Braaten}),
and we now have a remarkably complete characterization of these processes.

Our goal for this Letter is to examine four-body collisions in an ultracold two-component gas.  
In particular, we want to explore the possibility of producing 
Efimov trimers 
via four-body recombination.  Since the experimental evidence for the Efimov effect 
is from low-energy three-body collisions rather than from bound trimers, producing
the trimers is a natural next step.  One scheme for doing this in
the tight confinement of an optical lattice has been proposed~\cite{Kohler}, but here
we investigate one possibility for producing them in free space.

Besides the intrinsic interest of Efimov states, it is also important to
develop the theory of the fundamental process of four-body recombination quantum mechanically.
Compared with three-body collisions, our knowledge of ultracold four-body collisions is still 
quite rudimentary.  The reason is clear: solving the Schr\"odinger equation 
with three additional degrees of freedom is a much more difficult task.  
There have, of course, been many studies of four-body systems, but only recently
have some relevant to ultracold quantum gases begun to appear~\cite{Dimer_Dimer,Bcsbec_4b,HammerFourBoson,TomioFourBoson}
--- none of which have addressed four-body recombination.

The system we consider in this Letter is an ultracold mixture of atoms $A$ and 
$B$.  We take $A$ to be bosons, but all we need specify now about $B$ is that
they are distinguishable from $A$.  Our goal is to produce Efimov trimers $A_3$ via the 
four-body recombination process
\begin{equation}
A+A+A+B\longrightarrow A_3+B.
\label{Reaction}
\end{equation}
To this end, we will assume that the two-body $s$-wave scattering length among $A$ atoms $a_{AA}$
is infinite to give the most favorable case for Efimov states.  We will also assume that
the interspecies scattering length $a_{AB}$ is finite and that any dimer states, which are likely
to be present in real systems, lie much deeper than the weakly bound Efimov trimer states.

\begin{figure}
\includegraphics[width=50mm]{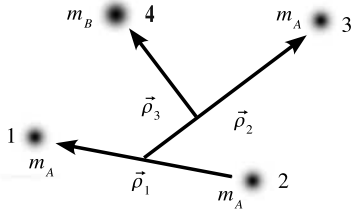}
\caption{The Jacobi coordinates for the four-body system. Atoms 1, 2, and 3 are identical bosons $A$
with mass $m_A$, while atom 4 is an atom of another species $B$ with mass $m_B$.}
\label{Fig:Jacobi}
\end{figure}

Our treatment of this process is asymmetrical in $A$ and $B$, mirroring the
differences in their relative scattering lengths.  
The two-body interactions for the $A_3$ subsystem are well-described by 
the zero-range model~\cite{Efimov}.
The advantage of this model is that the three-body solutions are known and especially simple
in the $|a_{AA}|\rightarrow \infty$ limit~\cite{Efimov}.  Note that extending this model to four-body
systems leads to equations that are not straightforward to solve.  Instead, we borrow an
idea from Rydberg physics~\cite{Greene_Ryd,Hamilton_Ryd} --- and from many-body theories of Bose-Einstein
condensates (BECs) as well~\cite{BECReview} --- to model the $AB$ interaction.
Since the wavelength of the $A$ atoms, either when they are free or in the Efimov molecule,
is much larger than the size of atom $B$, 
the $AB$ interactions
can be approximated by the Fermi contact potential~\cite{Fermi_Pot} (atomic units will be used throughout)
\begin{equation}
\label{Eq:Fermi_Pot}
V_{AB}(r_{i4})=\frac{2\pi a_{AB}}{\mu_{AB}}\delta(r_{i4}), \qquad i=1,2,3,
\end{equation}
where $\mu_{AB}$ is the two-body reduced mass and $r_{i4}$ is one of the $AB$ interparticle distances.
The Jacobi coordinates we use to represent the internal motion of the four particles are shown in Fig~\ref{Fig:Jacobi}.

Given the success of the adiabatic hyperspherical representation in describing the three-body
continuum~\cite{MacekRecomb,EsryRecomb}, we will use
it here to treat the four-body continuum as well.  Since we want to use
the known solutions for the $A_3$ subsystem, 
we build the four-body hyperspherical coordinates from the three-body ones~\cite{DelvesDef}.
The four-body hyperradius $R_4$ and hyperangle $\alpha_4$ are thus defined as
\begin{equation}
\label{Eq:R4}
\mu_4 R_4^2=\mu_{3,4}\rho_3^2+\mu_3 R_3^2 ~~~ {\rm and} ~~~ \tan \alpha_4=\sqrt{\frac{\mu_3}{\mu_{3,4}}}\frac{R_3}{\rho_3}.
\end{equation}
Here,
$\mu_3$=${m_{A}}/{\sqrt{3}}$, $\mu_{3,4}$=${3 m_A m_B}/(3 m_A\!\!+\!\!m_B)$, and
$\mu_4$=$\sqrt{\mu_3 \mu_{3,4}}$. 
Finally, $R_3$ denotes the three-body hyperradius; and $\alpha_3$, Delves' three-body hyperangle~\cite{DelvesDef}.

The representation we will use for the four-body problem, however, is not fully
adiabatic in that we will not include the interactions $V_{AB}$ in the
adiabatic Hamiltonian $H_{\rm ad}$.  Rather, we will include them later as coupling between
the channels.  This choice, together with our definition of coordinates, permits
separation of variables in the adiabatic equation, utilizing
the known solutions for the $A_3$ subsystem.
Mathematically, this procedure begins with the four-body Schr\"odinger equation 
\begin{equation*}
\label{Sch_Full}
\left[T_{R_4}\!+\!V_{AB}(r_{14})\!+\!V_{AB}(r_{24})\!+\!V_{AB}(r_{34})\!+\!H_{\mathrm{ad}}\right]\Psi=E\Psi,
\end{equation*}
where $T_{R_4}$ is the hyperradial kinetic energy and
\begin{equation}
\label{Sch_Ad}
H_{\mathrm{ad}}=T_{\Omega_4}+T_{\Omega_3}+V_{123},
\end{equation}
which includes the hyperangular kinetic energies $T_{\Omega_i}$
as well as all of the interactions among $A$ atoms in $V_{123}$.
The notation $\Omega_3$ denotes collectively all of the three-body
hyperangles; and $\Omega_4$, all remaining hyperangles for the four-body system.
In the ultracold limit, only the
zero total orbital angular momentum solution is relevant by the 
generalized Wigner threshold law~\cite{EsryThreshold}.  And, since
Efimov states only exist for zero orbital angular momentum of $A_3$,
the angular momentum of $B$ relative to the trimer must also be zero.
The channels for the four-body problem are thus defined from
\begin{equation}
H_{\rm ad} \Phi_\nu^{(4)} = U_\nu(R_4) \Phi_\nu^{(4)} .
\label{FourBodyAd}
\end{equation}
Separation of variables allows ($\nu\equiv\{\alpha,n\}$)
\begin{equation}
\Phi_\nu^{(4)}(R_4;\Omega_4,\Omega_3)=u_{\alpha n}(R_4;\Omega_4) \Phi_\alpha^{(3)}(\Omega_3).
\label{Phi4}
\end{equation}
Because we have chosen $a_{AA}\rightarrow\infty$, $\Phi_\alpha^{(3)}(\Omega_3)$ 
does not depend on $R_3$ --- otherwise this separation would only be approximate and
the resulting equations would have to be solved numerically.
For four-body recombination, we will need to find not only the 
$A_3+B$ bound channels, but also the four-body continuum channels $A+A+A+B$.

For the $A_3\!+\!B$ channels, we use ($\alpha$=0)~\cite{Efimov}
\begin{align}
\Phi_0^{(3)}(\Omega_3)&=\sum_{l=1}^3 \frac{2\sinh(s_0\alpha_3^{(l)})}{\sin(2\alpha_3^{(l)})} \label{Phi3} \\
(T_{\Omega_3}+V_{123})\Phi_0^{(3)}&=-\frac{s_0^2+\frac{1}{4}}{2\mu_3R_3^2}\Phi_0^{(3)}
\end{align}
where the summation is over the three possible three-body Jacobi sets,
each with its own Delves' hyperangle $\alpha_3^{(l)}$;
$s_0$$\approx$1.0062 is a universal constant.
Substituting Eq.~(\ref{Phi3}) into Eq.~(\ref{Phi4}), and the result into Eq.~(\ref{FourBodyAd}),
gives the 
equation for the bound trimer channels:
\begin{equation}
\left(\!-\frac{\partial^2}{\partial \alpha_4^2}\!-\!\frac{s_0^2+\frac{1}{4}}{\sin^2\!\alpha_4}\right)u_{0 n}=\lambda_{0 n}^2 u_{0 n},
~~ U_{0 n}=\frac{\lambda_{0 n}^2\!-\!\frac{1}{4}}{2\mu_4 R_4^2}.
\label{Eq:u}
\end{equation} 
The physically acceptable solution of Eq.~(\ref{Eq:u}) is
\begin{align}
u_{0 n}(\!R_4;\alpha_4)&=N_4\cos\alpha_4\sin^{\frac{1}{2}+i s_0}(\alpha_4)
\label{Eq:u_Soln}
\\
&\!\!\times{_2{\mathrm{F}}_1}(\frac{3}{4}\!+\!\frac{is_0}{2}\!\!-\!\frac{\lambda_{0 n}}{2},
\frac{3}{4}\!\!+\!\frac{is_0}{2}\!+\!\frac{\lambda_{0 n}}{2};\frac{3}{2};\cos^2\!\alpha_4) \nonumber
\end{align}
with $N_4$ the normalization constant.
To avoid the Thomas collapse~\cite{Thomas}, we take
the simple and expedient strategy of requiring that the three-body hyperradial wave functions 
vanish for $R_3$$\leqslant$$ R_0$, yielding a boundary condition on $u_{0 n}$:
$u_{0 n}(\alpha_4\!\!\leqslant \!\!\arcsin[\sqrt{\mu_3/\mu_4}R_0/R_4])$=0.
We note that none of our conclusions will depend on the
details of this regularization --- $R_0$ merely sets the scale for the features we predict.
Imposing this boundary condition leads to a transcendental equation for $\lambda_{0 n}$, and
it can be shown explicitly that the Efimov trimer energies are recovered from $U_{0 n}$ in the limit $R_4$$\rightarrow$$\infty$.
We include the diagonal coupling 
$Q_{\nu\nu}$=$\left<\!\!\left< \!\frac{d\Phi_\nu}{dR_4} \!\bigl|\! \frac{d\Phi_\nu}{dR_4} \!\right>\!\!\right>$ 
to get the most physical adiabatic potentials $W_\nu(R_4)\!=\!U_\nu(R_4)\!-\!\frac{1}{2\mu_4}Q_{\nu\nu}(R_4)$.
For $W_\nu(R_4)>0$ and $R_4\gg R_0$, 
\begin{equation}
{\small\!}W_\nu(R_4)\!\approx\!\!\left[\frac{2(\!n\!+\!\gamma)\!-\!s_0}{\pi}\!+\!\frac{1}{2}\!-\!
\frac{s_0}{\pi}\!\ln\!\left(\!
\frac{\mu_4 R_4^2}{\mu_3 R_0^2}\!\right)\right]\!\frac{1}{2\mu_4R_4^2} .
\label{Eq:Pot_Bound}
\end{equation}
Here, $\gamma$=0.30103 
and $n$=1,2,3,\ldots labels the trimer states.

We follow the same logic for the four-body continuum channels $A\!+\!A\!+\!A\!+\!B$.
The only difference is that $\Phi_\alpha^{(3)}$ is now a three-body continuum 
function, but still for $|a_{AA}|\rightarrow\infty$. That is, we replace $s_0$ in 
Eqs.~(\ref{Phi3})--(\ref{Eq:u_Soln}) by 
$i s_\alpha$ ($\alpha$$>$0), where $s_\alpha$ are real numbers determined by
\begin{equation}
\sqrt{3}s_\alpha\cos(\frac{\pi}{2}s_\alpha)=8\sin(\frac{\pi}{6}s_\alpha).
\label{3bodycont}
\end{equation}
The resulting $u_{\alpha m}$ satisfy the same boundary condition at $R_0$ as the bound channels.
Asymptotically, $\lambda_{\alpha m}\rightarrow s_{\alpha}\!+\!2m\!+\!\frac{3}{2}$ with $m$=0,1,2,\ldots
labeling the four-body continuum states possible for each $\alpha$.
The four-body continuum potentials thus behave as $1/R_4^2$ for $R_4\gg \!R_0$.

\begin{figure}
\includegraphics[width=75mm]{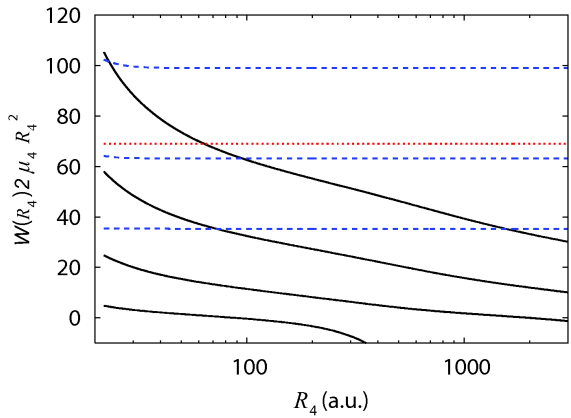}
\caption{(color online) 
The lowest four-body adiabatic hyperspherical potentials $W_\nu$ 
multiplied by $2\mu_4 R_4^2$ to better show their behavior. 
Black solid lines denote atom-trimer potentials for $\alpha$=0 and $n$=1,2,3,4; 
blue dashed lines, four-body continuum potentials for $\alpha$=1 ($s_1$=4.465) and $m$=0,1,2; 
and the red dotted line, the four-body continuum potential with $\alpha$=2 ($s_2$=6.818) and $m$=0. 
We take $R_0$=10~a.u. for all curves.}
\label{Fig:Pot}
\end{figure}
If we expand the total four-body wave function as 
\begin{equation}
\Psi(R_4,\Omega_4,\Omega_3) = \sum_\nu F_\nu(R_4) \Phi_\nu^{(4)}(R_4;\Omega_4,\Omega_3),
\end{equation}
then the non-adiabatic couplings
between four-body continuum and atom-trimer channels vanish since $\Phi_\alpha^{(3)}$ are independent of $R_3$
and form an orthonormal set.
Recombination is thus driven only by the diabatic couplings
\begin{equation}
V_{\nu'\nu}\!\!=\!\left<\nu'\lvert V_{AB}(\boldsymbol{r}_{14})\!+\!V_{AB}(\boldsymbol{r}_{24})\!+\!V_{AB}(\boldsymbol{r}_{34})\rvert \nu\right>
\end{equation}
and occurs predominantly at the crossings in Fig.~\ref{Fig:Pot}.
Figure~\ref{Fig:Pot} shows the lowest atom-trimer and four-body continuum potentials. 
Notice that the two sets of potentials cross in several places. The lowest two atom-trimer potentials, however, 
do not cross a continuum channel.  In our model, then, the lowest two Efimov trimers can only be populated
by weak non-adiabatic transitions between atom-trimer channels.  This conclusion is independent
of $R_0$, but for more realistic, finite range two-body potentials all the channels will likely be coupled
at small $R_4$.

Using Eq.~(\ref{Eq:Fermi_Pot}) and evaluating $V_{\nu\nu}$ numerically,
we find that it is proportional to $R_4^{-3}$ when $R_4\lesssim |a_{AB}|$.
But, four-body recombination should occur when $R_4$ is comparable to the size of the final three-body 
Efimov state. So, by making $|a_{AB}|$ small compared with
the size of the state we are interested in, we can neglect
$V_{\nu\nu}$. 
This requirement must be balanced against the need to have large off-diagonal
coupling elements $V_{\nu'\nu}$, although there will be transitions so long as
$a_{AB}$ is not zero.

In the adiabatic hyperspherical representation, four-body recombination Eq.~(\ref{Reaction}) 
starts on a four-body continuum channel at $R_4$$\rightarrow$$\infty$.  As the atoms
collide, they encounter an infinite number of
crossings with atom-trimer channels, one for each Efimov state, before reaching the classical
turning point.
Treating each crossing as independent from the others, we can estimate the transition probability
using the Landau-Zener approximation if $E$$>$$W_\nu(R_4^c)$, where 
$R_4^c$$\approx$$R_0\!\sqrt{\frac{\mu_3}{\mu_4}} \exp\!\left[\frac{\pi}{s_0}\!\left(n\!-\!\frac{1}{2}\lambda_\nu^c\!+\!
\frac{2\gamma-s_0}{2\pi}\!+\!\frac{1}{4}\right)\right]$ is the position of the crossing.
In this approximation, the recombination probability is given by
$P_{\mathrm{LZ}}=4T(1-T)\cos^2\!\Delta\phi$.  The relative phase $\Delta\phi$
is approximately zero since the potentials are nearly parallel near the crossings, and $T$ is
\begin{equation*}
T\!=
  \!\exp\!
   \left[ -
     \sqrt{\frac{\mu_3}{\mu_4}}\!
     \left(\!
      \frac{\mu_3}{\mu_{AB}}
      \frac{a_{AB}}{R_4^{c}}\!
     \right)^2\!\!\!
     \frac{\beta^2}{k_\nu^c R_0}
     \frac{2\pi}{(2\frac{s_0}{\pi}\lambda_\nu^c\!+\!\frac{1}{4})}
   \right]\!,
\end{equation*}
with $k_\nu^c$ the wavevector and
$\lambda_\nu^c$$\approx$$\sqrt{(s_\alpha\!+\!2m\!+\!\frac{3}{2})^2\!-\!\frac{1}{4}}$ 
the eigenvalue of Eq.~(\ref{Eq:u}) --- both evaluated at the crossing. 
The unitless constant $\beta$ originates from the evaluation of $V_{\nu'\nu}$,
has a weak dependence on the channel numbers, and is on the order of $10^{-3}$.
If $E<W_\nu(R_4^c)$, the system must tunnel in the initial continuum potential
to reach the crossing and make a transition.  Figure~\ref{Fig:Pot} and $R_4^c$ show that in this case,
though, there is always another energetically accessible crossing at larger
$R_4$ that will dominate the recombination.  
Consequently, we set $P_{\mathrm{LZ}}=0$ for all energetically closed crossings
since they do not alter the peak structure of the total recombination probability.
Further, we expect the first open
crossing beyond the classical turning point to dominate 
all other open crossings since the kinetic energy grows with $R_4$,
ensuring subsequent crossings will be traversed diabatically (i.e. without a transition).
For the same reason, we neglect the possibility of transitions from an atom-trimer
channel back to the four-body continuum at $R_4$$>$$R_4^c$.

\begin{figure}
\includegraphics[width=80mm]{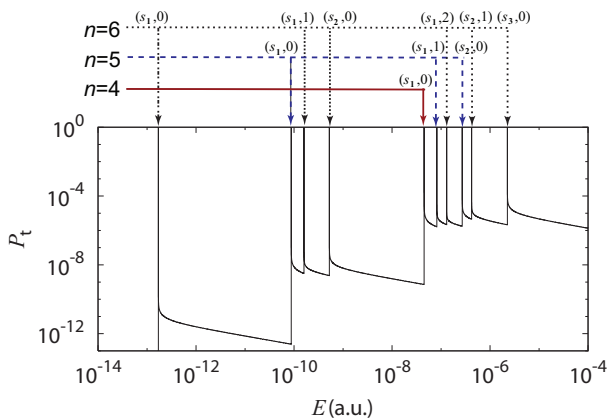}
\caption{The total recombination probability $P_T$ for atom-trimer channels up to $n$=6. 
$(s_\alpha,m)$ labels the incident channel that contributes to the indicated peak.
We have taken $R_0$=10~a.u. and $a_{AB}$=100~a.u.. 
}
\label{Fig:P}
\end{figure}

The four-body recombination rate $K_4$ is related to
$P_{\mathrm{LZ}}$ by $K_4\propto\frac{P_{\mathrm{LZ}}}{k^7}$ where $k=\sqrt{2\mu_4 E}$ 
is the incident four-body wave vector.
Because the couplings $V_{\nu'\nu}$ are quite small, 
the peaks of $P_{\mathrm{LZ}}$ occur at $E$$\approx$$W_\nu(R_4^c)$,
\begin{equation*}
W_\nu(R_4^c)\approx \frac{\left(s_\alpha\!+\!2m\!+\!\frac{3}{2}\right)^2}{2\mu_{3,4}R_0^2}
   e^{-\frac{\pi}{s_0}\!\left(2n-s_\alpha\!-2m+\frac{2\gamma-s_0}{\pi}-\!1\right)}.
\end{equation*}
This expression shows that for a given initial channel $(s_\alpha,m)$ there is a geometrically
spaced sequence of peaks in energy for recombination to Efimov trimers. This characteristic
feature of the three-body Efimov physics thus comes through in the four-body physics as well.  Moreover,
the spacing of the features is $\exp(\!-2\pi/s_0)$ just as one would predict from the three-body
physics.  To illustrate
this point, we show in Fig.~\ref{Fig:P} the total recombination probability $P_T$ calculated
as a sum of $P_{\rm LZ}$ over all initial and final channels.  Note that the width of the peaks
increases with $|a_{AB}|$ as does their shift from $W_\nu(R_4^c)$.

Up to this point, our analysis has assumed $|a_{\rm AA}|=\infty$.
If we let it be finite instead, then the above analysis applies in the regime 
$|a_{AA}|\gg|a_{AB}|$. To avoid the complication
of a $A^*_2+A+B$ continuum degenerate with the channels of interest,
we further require $a_{AA}$$<$0 so that there are no weakly bound dimers. Under these
conditions, the four-body adiabatic potentials behave as described above in the region
$|a_{AB}|\ll R_4\ll \lvert a_{AA}\rvert$.  For $R_4 \gg |a_{AA}|$,
the four-body continuum potentials approach the four-body hyperspherical harmonic potentials
$W_\nu\approx\frac{\lambda(\lambda+7)+12}{2\mu_4 R_4^2}$ with $\lambda$ a non-negative integer, and 
the atom-trimer potentials approach the trimer bound energies $W_\nu\approx E_{n}$. 
It follows that at energies 
$(2\mu_4 a_{\rm AA}^2)^{-1}$$\ll$$ E$$\ll$$(2\mu_4 a_{\rm AB}^2)^{-1}$ $K_4$ keeps the structure described above.

It is in the zero energy limit that the effects of finite $a_{\mathrm{AA}}$ reveal themselves, and
we will use WKB to explore this limit~\cite{Jose_Scale}.
When $E\rightarrow 0$, recombination into the most weakly bound Efimov trimer dominates. Since the size of 
this trimer is on the order of $|a_{\rm AA}|$, we expect recombination to occur at $R_4$$\approx$$ |a_{\rm AA}|$.
So, following an analysis similar to that described in Ref.~\cite{Jose_Scale}, we find that 
the recombination probability is
\begin{equation}
P\propto (k|a_{AA}|)^7 \sin^2(k_n\lvert a_{\mathrm{AA}}\rvert+\Phi), 
\label{Finite_a}
\end{equation}
where $\Phi$ is a short-range phase independent of $a_{\mathrm{AA}}$ and
$k_n$=$\frac{2}{R_0}\frac{\mu_4}{\mu_3}\exp[-(n\pi\!+\!\gamma)/s_0]$
is the wave number for the final trimer state $n$ at $E$=0.
When $\lvert a_{\mathrm{AA}}\rvert$ increases by a factor of 22.7, a new
atom-trimer channel appears and $k_n$ changes to $k_{n+1}$ which, in turn,
changes the period of the $a_{\mathrm{AA}}$-dependent oscillations.
It turns out that recombination into a particular atom-trimer state will show about
seven full oscillations in $K_4$ as a function of $a_{AA}$.  From the
relation $K_4$$\propto$$P/k^7$, Eq.~(\ref{Finite_a}) also shows that $K_4$
will be constant in the threshold regime, $E$$\lesssim$$(2\mu_4 a_{AA}^2)^{-1}$.

If the trimers cannot be experimentally observed directly, then their
production can be monitored through the loss of either $A$ or $B$ atoms.
In an ultracold mixture of the two, it is much better to monitor the $B$ atoms, however,
as other few-body processes will lead to loss of $A$ atoms and thus mask the effects
we predict. 
The best scenario makes
the $B$ atoms spin-polarized fermions. In this case, $A$+$A$+$A$+$B$ recombination is unaffected,
but the competing loss processes involving two or more $B$ atoms can be completely avoided as they will be suppressed
near threshold~\cite{EsryThreshold}.  There remains the possibility for loss of $B$ atoms
in $A$+$A$+$B$ collisions, but for
$|a_{AA}|$=$\infty$ or $a_{AA}$ finite but negative, the three-body potentials 
are all repulsive~\cite{Jose_OneRes} and only deeply-bound dimer channels are available. 
The various loss rates for this system are thus small~\cite{Jose_OneRes}. 
Monitoring the loss of $B$ atoms should then provide a signature of Efimov trimer formation.

To summarize, we have taken the first steps in understanding ultracold four-body
recombination into Efimov trimers.  By carefully setting up the problem, we
were able to obtain largely analtyical results.  In the process, we showed
that the four-body recombination should show prominent, geometrically spaced peaks that reflect
the three-body Efimov physics.  In fact, these peaks are separated by precisely
the factor one expects from the three-body physics.  We also suggested one
approach for experimentally determining whether four-body recombination was
indeed occurring.  Should an experiment reach this milestone, then our work
further suggests the interesting possibility that one might be able to use the temperature to control which
Efimov state is produced.

\begin{acknowledgments}
We are grateful to J.P. D'Incao for early discussions of this work.  This work was
supported in part by the National Science Foundation and in part by the
Air Force Office of Scientific Research.
\end{acknowledgments}

\end{document}